\documentclass[aip,amsmath,graphicx]{revtex4-1}
\pdfoutput=1
\usepackage{graphicx}
\usepackage{dcolumn}
\usepackage{bm}

\usepackage[utf8]{inputenc}
\usepackage[T1]{fontenc}
\usepackage{mathptmx}
\usepackage{etoolbox}
\usepackage[table,xcdraw]{xcolor}

\draft 

\begin{document}


\title{Optimized higher-order photon state classification by
machine learning} 



\author{Guangpeng Xu}
\author{Jeffrey Carvalho}
\author{Chiran Wijesundara}
\author{Tim Thomay}
 \email{thomay@buffalo.edu.}
\affiliation{Department of Physics, University at Buffalo, State University of New York, Buffalo, NY, 14260, USA
}%


\date{\today}

\begin{abstract}
The classification of higher-order photon emission becomes important with more methods being developed for deterministic multiphoton generation. 
The widely-used second-order correlation \(g^{(2)}\) is not sufficient to determine the quantum purity of higher photon Fock states.
Traditional characterization methods require a large amount of photon detection events which leads to increased measurement and computation time. 
Here, we demonstrate a Machine Learning model based on a 2D Convolutional Neural Network (CNN) for rapid classification of multiphoton Fock states up to \(|3\rangle\) with an overall accuracy of 94\%. 
By fitting the \(g^{(3)}\) correlation with simulated photon detection events, the model exhibits efficient performance particularly with sparse correlation data, with 800 co-detection events to achieve an accuracy of 90\%. 
Using the proposed experimental setup, this CNN classifier opens up the possibility for quasi real-time classification of higher photon states, which holds broad applications in quantum technologies.
\end{abstract}


\maketitle 

\section{Introduction}
Recently, quantum light has been undergoing rapid advancements and playing a pivotal role in the development of quantum technologies
\cite{zhang_quantum_2017, yang_efficient_2016, mittal_topological_2018, he_carbon_2018, hu_experimental_2016}. 
At the quantum level, the photon nature of light is characterized as discrete packets of energy \cite{fox_quantum_2006,einstein_uber_1905a}, offering remarkable precision, sensitivity, and enhanced communication security beyond what classical optics can achieve \cite{obrien_photonic_2009,wang_integrated_2020,lodahl_chiral_2017} and thereby facilitating the application of quantum systems in diverse fields such as metrology \cite{ye_quantum_2008,pinel_ultimate_2012,dowling_quantum_2015,polino_photonic_2020}, computer science \cite{walmsley_light_2023,ladd_quantum_2010,walther_experimental_2005}, and communication \cite{ursin_entanglementbased_2007,duan_longdistance_2001,yuan_entangled_2010,pan_entanglement_2001}.

However, due to the typically low emission rate and detection inefficiency, experimental observation on quantum emitters often necessitates a lengthy experiment time and produces large datasets with a high level of noise \cite{kuhn_deterministic_2002, hennrich_photon_2004, moreau_singlemode_2001}, which renders the fitting process computationally expensive \cite{lin_efficient_2021, kudyshev_machine_2021}.
The identification of multiphoton states is commonly addressed by assembling multiple single-photon detectors that are time-correlated \cite{hockel_direct_2011, elvira_higherorder_2011, stevens_thirdorder_2014, steudle_measuring_2012,jonsson_evaluating_2019}.
This exacerbates the challenge of data analysis, as traditional computational techniques like the Levenberg-Marquardt (L-M) Method \cite{marquardt_algorithm_1963} require extensive co-detection events to achieve a satisfactory accuracy \cite{kudyshev_machine_2021}.

Over the past decade, novel data-driven formalisms such as Machine Learning (ML) have introduced new possibilities in quantum photonics experiments \cite{yao_intelligent_2019a, zhou_emerging_2019, kudyshev_machine_2021a, yu_reconstruction_2019, wang_experimental_2017}.
Specialized in analyzing large and sparse datasets, ML models have provided speedup by orders of magnitude in certain quantum measurements \cite{hegde_deep_2020, freire_artificial_2023}, and show potential in overcoming the inherent limitations of conventional fitting methods particularly in the low-photon flux regime \cite{wu_classification_2023}.
For example, a Convolutional Neural Network (CNN)-based algorithm was developed for rapid classification of single photon emitters in the NV center of nanodiamonds \cite{kudyshev_rapid_2020}.
Compared to the L-M method, the accuracy is improved with the CNN model by recognizing subtle features extracted from sparse correlation data.
A single artificial neuron model was developed to reduce the required average number of photons down to less than one for distinguishing thermal light from coherent light in low-light measurements \cite{you_identification_2020}. 
A study by Cortes et al \cite{cortes_accelerating_2020} demonstrated that employing statistical learning methods for the reconstruction of \(g^{(2)}\) data can substantially accelerate the data acquisition process from few-shot measurements.

While considerable efforts were directed towards single-photon emitters \cite{khalid_perfect_2024, eisaman_invited_2011, aharonovich_solidstate_2016, palacios-berraquero_largescale_2017}, the emission of multiple, indistinguishable photons also becomes favorable for quantum systems \cite{dellanno_multiphoton_2006, laiho_measuring_2022}, making them promising candidates to exert further influence on various quantum applications such as Boson sampling \cite{wang_boson_2019}.
As the commonly adopted \(g^{(2)}\) correlation proves inadequate when detecting the photon "superbunching" in higher Fock states, it necessitates the introduction of higher-order correlation \cite{laiho_measuring_2022, grunwald_nonquantum_2020, zubizarretacasalengua_structure_2017}.

In this study, we present a 2D CNN based ML model for rapid classification of multiphoton states, including photon Fock states up to \(|3\rangle\) and coherent states of laser emission.
The time-dependent photon detection data are simulated, and by mixing each Fock state with the corresponding coherent state, the quantum purity of emitters can be manipulated.
\(g^{(3)}\) correlation is performed on the simulation data and fitted using a supervised machine learning model to return the photon state classification results.
Through model training and optimization, the average accuracy of classification surpasses 90\% for all the Fock states, with an overall accuracy of 94\%.
The model exhibits effective performance with sparse datasets, with only 800 photon detection events to achieve a 90\% average accuracy.
Finally, we propose an experimental setup for quasi-real time photon state classification accelerated by the ML model.
For the first time, a 2D CNN algorithm is employed for identifying multiphoton states and shows enhanced accuracy and data efficiency.


%
%
\section{Methods}
\begin{figure}[!t]
    \includegraphics{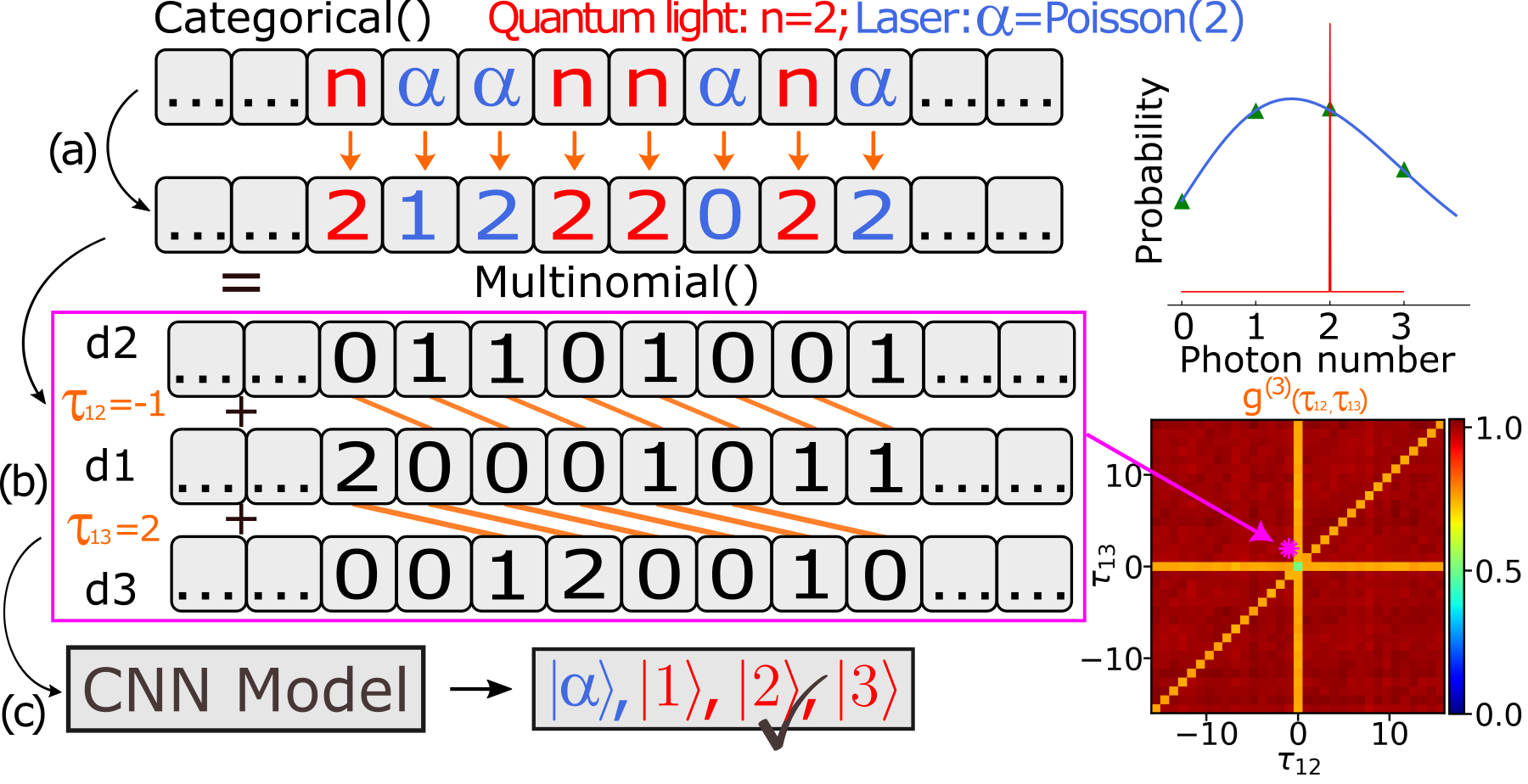}
    \caption{
    Simulation of HBT experiment, $g^{(3)}$ correlation algorithm and rapid Fock state classification based on Machine Learning.  
    (a) Schematic view of Monte-Carlo simulation. The categorical function produces a list of label markers distinguishing between quantum light and laser, where in this instance, "n" represents a photon at Fock state $|2\rangle$, while "$\alpha$" means a coherent state $|\alpha\rangle$ with average number of photons set to 2. The portion of quantum light labels and laser labels in the list is determined by the quantum light probability, set here at 50\%, indicating an equal mixture of both. Each marker is then replaced by the actual photon number on the second row, depending on its distribution: each quantum light "n" is replaced by 2, from its delta distribution shown by the red plot on the right; while each "$\alpha$" is replaced with an integer generated by a Poissonian distribution function with an average of 2, as illustrated by the blue plot.
    (b) $g^{(3)}$ correlation algorithm. In the purple rectangle, the second row with photon numbers is equally split into three rows using a multinomial function, representing photon events detected by three virtual detectors labeled as 'd1', 'd2' and 'd3'. The yellow lines linking detection events from different detectors illustrate the algorithm of $g^{(3)}$ correlation associated with a particular data point marked in purple on the $g^{(3)}$ heat map to the right. Here,  $\tau_{12}$ and  $\tau_{13}$, representing the time differences between detector 1 and 2, and detector 1 and 3, are -1 and 2, respectively.
    (c) CNN-based Fock state classification. The $g^{(3)}$ correlation results are fed into a CNN model that is pre-trained with similar simulated data, and the model yields the classification result.
    }
\end{figure}
To simulate photon correlation experiment within an extended Hanbury Brown and Twiss (HBT) scheme \cite{brown_correlation_1956} shown in Fig.7, the Monte-Carlo Method is used to generate photon detection events with the arrival timestamps.
The simulation model consists of three primary parts: photon stream emission, transport, and detection as the output.
This model is built upon the TensorFlow Probability (TFP) extension and the TensorFlow Distribution (TFD) probabilistic model \cite{abadi_tensorflow_}.

To emulate various quantum pureness of emitters and experimental factors that cause deviation on \(g^{(k)}(0)\), we incorporate each Fock state \(|n\rangle\) with a corresponding coherent state \(|\alpha\rangle\), with both states sharing the same average photon number.
The expectation of photon number for a coherent state is given by \(\langle\alpha|\hat{n}|\alpha\rangle={|\alpha|}^{2}\), hence \(\alpha=\sqrt{\overline{n}}\). \cite{loudon_quantum_2000}
A pure coherent state with a temporally random sequence of photons produces the correlation \(g^{(k)}(0)\) to always be 1, implying no correlation between the detection events \cite{fox_quantum_2006}.
When mixing a Fock state and the corresponding coherent state, the variance of the photon number is increased while the average photon number remains the same.
Hence the photon distribution of such mixture lies between a delta function characterizing the Fock state, and a Poissonian distribution for the coherent state.
And the \(g^{(k)}(0)\) values of the superposition state is calculated by Eqs. (1):

\begin{subequations}
\label{eq:whole}
\begin{equation}
g^{(k)}(0) = \frac{1}{n^k} \frac{n!}{(n-k)!}p+(1-p), \;\;  n\geq k\label{subeq:1}
\end{equation}
\begin{equation}
g^{(k)}(0) = 1-p, \;\;  n<k\label{subeq:2}
\end{equation}
\end{subequations}

where $n$ means the photon Fock state, $p$ represents the probability of the Fock state in the mixture, and $(1-p)$ denotes the portion of the coherent state.
By manipulating the contribution of each state, the lower or upper bound of \(g^{(k)}(0)\) can be reached, representing an ideal quantum emitter at state \(|n\rangle\) or a coherent source at state \(|\alpha\rangle\) with \(\alpha=\sqrt{n}\), respectively.

Fig.1 illustrates the simulation of an imperfect $|2\rangle$ state emitter as an example.
To simulate the photon emission, the TFD Categorical function is used to generate a list of light source labels, shown by the top row in Fig.1 (a).
Each label represents either a quantum emission in the \(|n\rangle\) state (red), or coherent laser emission \(|\alpha\rangle\) (blue) with an average photon number of \(n\).
While for an ideal quantum emitter no laser labels will be included, non-ideal quantum emitters contain a mixture of both quantum emission labels and laser labels that are randomly distributed within the list.
The quality of quantum emitters can be assessed by the portion of quantum emission labels in the list, controlled by a simulation parameter called "quantum light probability" (QLP) that ranges from 0 to 1.
In practical experiments, the significance of QLP reflects a cumulative effect from various factors such as the background signal of classical light, that impact the correlation result $g^{(k)}(0)$.
Shown in Fig. 1, the \(|2\rangle\) state emitter with a QLP of 0.5 results in an even distribution of half quantum emission labels and half laser labels.
Each light label is then replaced by an integers to represent the number of photons in each emission, illustrated by the second row in Fig.1 (a).
For quantum emission labels, it's straightforward to fill in an integer 2 (in red) to represent the emission of two identical photons from the \(|2\rangle\) emitter.
The laser labels are substituted by using the TFD Poisson function, which generates integers (in blue) in a Poissonian distribution with the average value of 2. 
This adjustment comes from the photon number for each emission from a coherent laser follows a Poissonian distribution shown by the blue plot on the right of Fig.1 (a), instead of being a constant for quantum emission represented by the red plot.

With the substitution of photon numbers, the emulated photon stream is fed into a virtual multi-port beam splitter using the TFD multinomial function.
Photons in each emission packet are split into three paths with multinomial distribution, as shown in the purple rectangle in Fig.1 (b).
The split photon events are recorded by virtual photon detectors (noted by "d1", "d2" and "d3" in Fig.1 (b)) and output as simulation results, including the number of photons detected each time, along with the corresponding timestamp.
For convenience and consistency, we assume all light emitters are periodically excited by a pulsed source, with the pulse period serving as the unit of time throughout this simulation.
For simplicity, the photon number of each detection can be resolved by the virtual detectors.
Experimental imperfections such as optical loss, dark counts and detection inefficiency are not directly simulated.
As these imperfections can be addressed either through post-selections with respect to the triggering signal or through the use of low-loss photon detectors \cite{schapeler_how_2023,lita_counting_2008}, without significantly altering the overall photon distribution.
The total number of detection events is another parameter that represents the experiment time and determines the data sparsity of \(g^{(3)}\) correlation.

\begin{equation}
g^{(3)}(\tau_{12},\tau_{13}) = 
\frac{
\langle n_1(t) * n_2(t+\tau_{12}) * n_3(t+\tau_{13}) \rangle 
}{
\langle n_1 \rangle * \langle n_2 \rangle  * \langle n_3 \rangle 
}
\end{equation}

The detected photon events are analyzed using the \(g^{(3)}\) function, by Eq. (2). 
The time differences \(\tau_{12}\), and \(\tau_{13}\) between detector 1 and 2, and detector 1 and 3, respectively, are variables to compute the normalized \(g^{(3)}(\tau_{12},\tau_{13})\) correlation, shown by the heat map on the right of Fig.1 (b).
The data point at (\(\tau_{12}=-1, \tau_{13}=2)\) marked in purple on the heat map is chosen to demonstrate the \(g^{(3)}\) algorithm depicted with the yellow stripes in Fig.1 (b).
With these specific time differences, the d2 vector is shifted forward by 1 unit relative to d1, while the d3 vector is shifted backward by 2 units.
The dot product of the three vectors is divided by the vector length and normalized by its mean, shown by the purple-marked pixel in Fig.1 (b).
The \(g^{(3)}\) function implements the above algorithm for all \(\tau_{12}\) and \(\tau_{13}\) variables, ranging from -16 to 16 with an interval of 1, and returns a two-dimensional matrix of \(g^{(3)}\) correlation results.
A 2D CNN model is developed for photon state classification based on the correlation results, shown in Fig.1 (c).
In the given example, the \(|2\rangle\) state quantum emitter is successfully identified among four options: laser within a coherent state and quantum emitters in \(|1\rangle\), \(|2\rangle\), and \(|3\rangle\) states, respectively.

According to Eqs. (1), to characterize a multiphoton Fock state $|n\rangle$, a k-th order correlation with $k>n$ is required to ensure the central point $g^{(n)}(0)$ remains zero.
Although the presented algorithm is extensible for high-order correlations, due to the rapid scaling of computational complexity with the order of correlation \cite{kudyshev_machine_2021}, $g^{(3)}$ is chosen for categorizing and characterizing multiphoton states within a manageable computation time.
Taking the time difference $\tau$ as the variable for $g^{(2)}$, the $g^{(3)}$ function has two variables \(\tau_{12}\), and \(\tau_{13}\) represented by x and y axes, being the time differences between detector 1 and 2, and detector 1 and 3, respectively.
The third time difference \(\tau_{23}\), which can be recursively derived from the other two as \(\tau_{23} = \tau_{13}-\tau_{12}\), is not considered as a variable in this context.

\begin{table}[]
\begin{tabular}{|c|cc|cc|cc|}
\hline
 &
  \multicolumn{2}{c|}{\(|1\rangle\)} &
  \multicolumn{2}{c|}{\(|2\rangle\)} &
  \multicolumn{2}{c|}{\(|3\rangle\)} \\ \hline
\begin{tabular}[c]{@{}c@{}}Quantum Light\\  Probability\end{tabular} &
  \multicolumn{1}{c|}{\(g^{(2)}(0)\)} &
  \(g^{(3)}(0)\) &
  \multicolumn{1}{c|}{\(g^{(2)}(0)\)} &
  \(g^{(3)}(0)\) &
  \multicolumn{1}{c|}{\(g^{(2)}(0)\)} &
  \(g^{(3)}(0)\) \\ \hline
{\color[HTML]{3531FF} 0.00} &
  \multicolumn{1}{c|}{{\color[HTML]{3531FF} 1.00}} &
  {\color[HTML]{3531FF} 1.00} &
  \multicolumn{1}{c|}{{\color[HTML]{3531FF} 1.00}} &
  {\color[HTML]{3531FF} 1.00} &
  \multicolumn{1}{c|}{{\color[HTML]{3531FF} 1.00}} &
  {\color[HTML]{3531FF} 1.00} \\ \hline
{\color[HTML]{3531FF} 0.25} &
  \multicolumn{1}{c|}{{\color[HTML]{3531FF} 0.75}} &
  {\color[HTML]{3531FF} 0.75} &
  \multicolumn{1}{c|}{{\color[HTML]{3531FF} 0.88}} &
  {\color[HTML]{3531FF} 0.75} &
  \multicolumn{1}{c|}{{\color[HTML]{3531FF} 0.92}} &
  {\color[HTML]{3531FF} 0.81} \\ \hline
{\color[HTML]{036400} 0.50} &
  \multicolumn{1}{c|}{{\color[HTML]{036400} 0.50}} &
  {\color[HTML]{036400} 0.50} &
  \multicolumn{1}{c|}{{\color[HTML]{036400} 0.75}} &
  {\color[HTML]{036400} 0.50} &
  \multicolumn{1}{c|}{{\color[HTML]{036400} 0.83}} &
  {\color[HTML]{036400} 0.61} \\ \hline
{\color[HTML]{CB0000} 0.75} &
  \multicolumn{1}{c|}{{\color[HTML]{CB0000} 0.25}} &
  {\color[HTML]{CB0000} 0.25} &
  \multicolumn{1}{c|}{{\color[HTML]{CB0000} 0.62}} &
  {\color[HTML]{CB0000} 0.25} &
  \multicolumn{1}{c|}{{\color[HTML]{CB0000} 0.75}} &
  {\color[HTML]{CB0000} 0.42} \\ \hline
{\color[HTML]{CB0000} 1.00} &
  \multicolumn{1}{c|}{{\color[HTML]{CB0000} 0.00}} &
  {\color[HTML]{CB0000} 0.00} &
  \multicolumn{1}{c|}{{\color[HTML]{CB0000} 0.50}} &
  {\color[HTML]{CB0000} 0.00} &
  \multicolumn{1}{c|}{{\color[HTML]{CB0000} 0.67}} &
  {\color[HTML]{CB0000} 0.22} \\ \hline
\end{tabular}
\caption{
Theoretical $g^{(2)}$ and $g^{(3)}$ values of Fock states when mixing with the corresponding coherent states. The quantum light probability indicates the portion of quantum light in the simulation when it is mixed with a corresponding coherent state. A probability of 0.5 (in green) signifies a light source composed of equal parts quantum light and laser, which is set to be the critical point distinguishing between a quantum emitter and a coherent laser. Values below 0.5 (in blue) or above 0.5 (in red) are labeled as a laser or quantum light, respectively.
}
\label{table:1}
\end{table}

Table 1 provides theoretical $g^{(2)}(0)$ and $g^{(3)}(0)$ values for Fock states $|1\rangle$, $|2\rangle$ and $|3\rangle$ when mixed with corresponding coherent states $|\alpha\rangle$.
The quantum light probability (QLP) represents the portion of Fock states in the mixture, e.g., $QLP=1$ denotes an ideal quantum state, while $QLP=0$ is for a coherent state.
During the simulation, coherent states with an average photon number of three or less are modeled, which is often achieved by strongly attenuating a classical source, such as a laser, when measuring quantum emission \cite{thomay_simultaneous_2017a}.
The $g^{(3)}$ results in Fig.2 align well with the $g^{(3)}_{|2\rangle}$ values in Table 1:
Firstly, the zero-delay value $g^{(3)}_{|2\rangle}(0)$ is 0, as the order of correlation 3 is greater than the value of the Fock state 2.
Secondly, the data within the cross pattern and the anti-diagonal that are in green, approaches 0.5 due to the special values of time differences.
For the vertical (or horizontal) stroke of the cross, the time difference \(\tau_{12}\) (or \(\tau_{13}\)) is 0. Meanwhile, for the anti-diagonal (bottom left to top right), the two time differences are equal, signifying that \(\tau_{13}\) is 0.
With one time difference being zero, the $g^{(3)}$ correlation can be treated to be equivalent to a $g^{(2)}$, which correlates only two detectors with one variable of time difference.
Hence the value of 0.5 on the heatmap can be explained with $g^{(2)}_{|2\rangle}(0) = 0.5$.
Finally, the remaining data points are normalized with respect to the average photon number, typically resulting in a normalization factor of 1.
To better visualize the 2D $g^{(3)}$ matrix, three cross-sections marked with colored arrowheads on the heatmap are plotted bar charts on the right side of Fig.2.
Each represents the evolution of the $g^{(3)}$ as a function of only \(\tau_{12}\)), with \(\tau_{13}\)) being set at special values: \(\tau_{13}=-\tau_{12}\) for black bars, \(\tau_{13}=0\) for purple, and \(\tau_{13}=\tau_{12}\) for yellow, corresponding to the main diagonal, horizontal stroke in the cross pattern, and the anti-diagonal, on the heatmap.

\begin{figure}[!t]
    \includegraphics{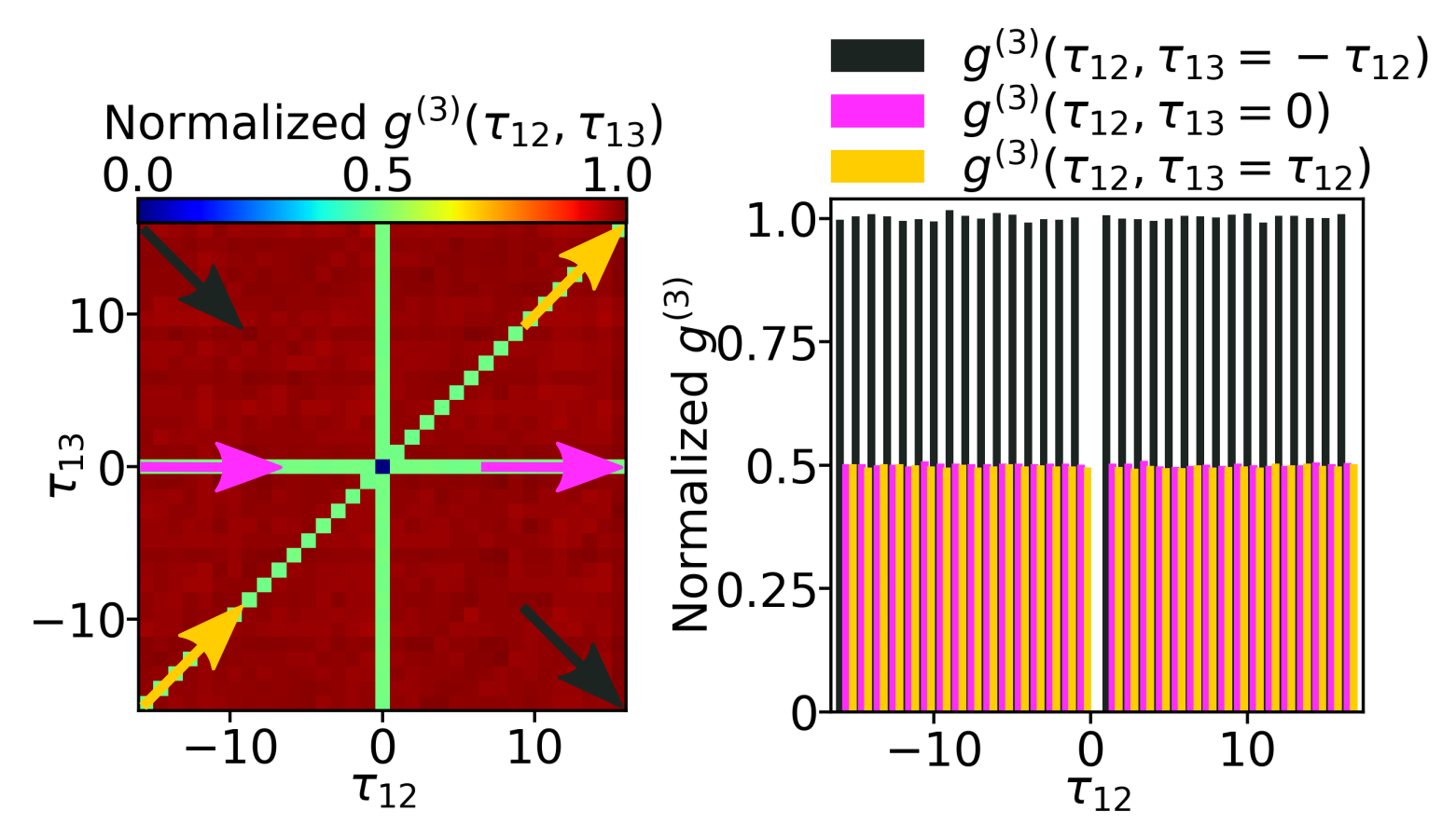}
    \caption{
    An example of $g^{(3)}$ correlation results from simulated data for a $|2\rangle$ Fock state without mixing with a coherent state.   
    Left: Normalized $g^{(3)}$ correlation in a heat map, as a function of $\tau_{12}$ and $\tau_{13}$, which represent the time differences between detector 1 and 2, and detector 1 and 3, respectively.   
    Right: 2D bar plots of $g^{(3)}$ correlation, as a function of only $\tau_{12}$, while $\tau_{13}$ is set to be either negative $\tau_{12}$, 0 or equal to $\tau_{12}$, corresponding to black, magenta and yellow bars. The trace of each bar plot is indicated by the arrows within the same color on the heat map.
    }
\end{figure}

In contrast to the second-order correlation, which has only one critical point $g^{(2)}(0)$ for categorization, the third-order correlation $g^{(3)}$ features multiple critical points: one at $g^{(3)}(0)$, and multiple at $g^{(3)}(0,\tau\neq0)$, effectively serving as $g^{(2)}(0)$.
In accordance with Eqs. (1), the complete $g^{(3)}$ cross correlation enables the identification of unknown Fock state emission and the assessment of its quantum purity.
Even for Fock states higher than \(|2\rangle\), although the $g^{(3)}(0)$ is no longer 0, it can still be determined from the $g^{(3)}(0)$ and $g^{(2)}(0)$ values.
Meanwhile, the dynamics of excitons can be further explored through other correlation algorithms, such as two-time second-order autocorrelation \cite{flagg_dynamics_2012a}.
Additionally, the two-dimensional nature of the $g^{(3)}$ data proves advantageous for CNN models that are adept at extracting spatial features from images \cite{egmont-petersen_image_2002, lecun_convolutional_2010}.

\begin{figure}[!t]
    \includegraphics{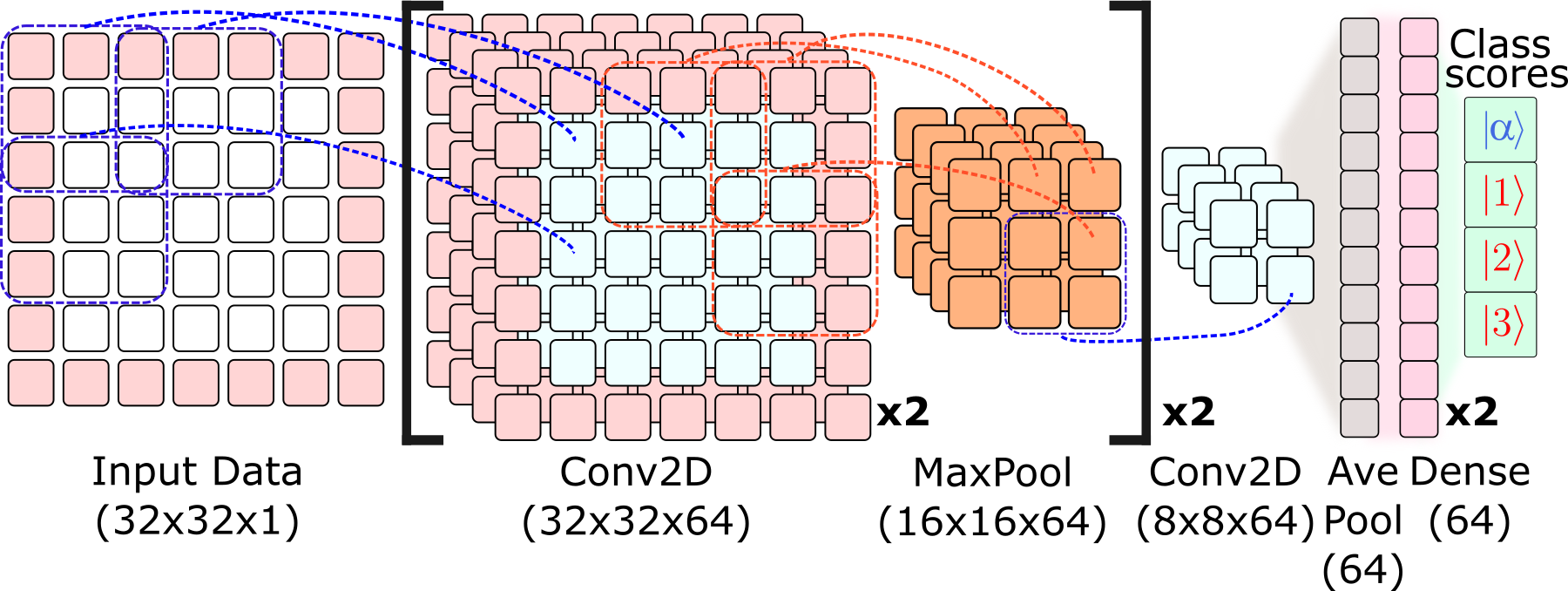}
    \caption{
    Schematics of the CNN model. Conv2D is for 2D convolution layer, MaxPool is for 2D max pooling layer, AvePool is for 2D global average pooling layer and x2 represents a layer or sequence of operations (in a bracket) that is repeated twice. The shape of output data from each layer is written in the parentheses. Dashed lines illustrate the connection between input data points and respective output data during layer operations. The initial input data consists of g3 correlation results in a 2D matrix, shown by the white cells, while red cells represent zero paddings. The Conv2D layer produces 3D matrices of data with a depth of 64, equal to the number of convolutional kernels (blue dashed lines represent one of the kernels). The data shape is condensed to one dimension at the AvePool layer, and the final dense layer outputs a score vector, indicating the confidence of predictions for each light source category.
    }
\end{figure}

While the \(g^{(3)}\) critical values can typically be determined using normal fitting methods, the experimental data often exhibits significant sparsity due to the limited number of photon detection events, entailing the introduction of advanced data fitting techniques.
To efficiently categorize photon states by fitting the \(g^{(3)}\) data, a machine learning model is developed based on the open-source API Keras \cite{kerasteam_2024}.
The model architecture and layer operations are illustrated in Fig.3. 
The model mainly comprises 2D convolution layers (Conv2D), 2D max pooling layers (MaxPool), 2D global average pooling layers (AvePool) and dense layers.
The output shape of each layer is noted in parenthesis.

Before passing to any layers, the correlation results are truncated and rescaled to improve the model's performance. 
The first 32 elements in each dimension of the original 33x33 \(g^{(3)}\) matrix are retained so that the truncated matrix, with a shape of 32x32, aligns with the preference of CNN layers for input data dimensions that are multiples of 2.
In cases of simulation with few-shot data, where normalized \(g^{(3)}\) results may contain extremely high values, the rescaling ensures all the \(g^{(3)}\) elements fall within the range of 0 to 1. 
The preprocessed datasets, depicted by the white square cells in the first diagram of Fig.3, have a spatial dimension of 32x32 and a depth of 1.

As the core of the CNN model, a 2D convolutional layer connecting to the input layer extracts the spatial features.
The input data is zero-padded on the outer border, as indicated by the red square cells in Fig.3, to prevent rapid degradation of the information at original borders, and maintain the output dimension of the Conv2D layer at 32x32.
Each Conv2D layer contains 64 convolutional kernels and each kernel is a 3x3 matrix that spatially slides across the data array with a step of 1.
The blue dashed boxes in Fig.3 illustrate the sum product calculation of a convolutional kernel, with the dashed lines connecting to the cells representing the results.
The above operation is performed by all the 64 kernels, with each kernel capturing an unique feature from the input, producing the Conv2D output with a depth of 64.
The zero-padding is also applied to the Conv2D output data to maintain the layer dimension.
Two of Conv2D layers are sequentially stacked and followed by a 2D max pooling layer, which calculates the maximum values of each pooling patch to spatially downsample the data dimension.
Three patches are depicted by the red dashed boxes in Fig.3 for visualization, and this process is repeated for all the 64 slices in depth.
The layer dimension is halved with a step size of 2 for max pooling, resulting in an output shape of 16x16x64. 
The number of trainable parameters is significantly reduced by the downsampling, so that a relatively low model capacity is maintained without losing essential information \cite{nagi_maxpooling_2011}.
The architecture enclosed in the square bracket in Fig.3, composed of two consecutive convolutional layers and one MaxPooling layer, is executed twice for learning hierarchical features.

The following global average pooling layer computes the average for each slice in depth, shown by the grey cells on the figure.
Each average reflects the importance of an extracted feature from the neural network.
This array of feature scores is analyzed by two dense layers, with each neuron fully connected to all the neurons from the previous layer.
The output layer computes scores for four potential photon states: the coherent state, as well as Fock states \(|1\rangle\), \(|2\rangle\) and \(|3\rangle\).
Each score indicates the model's confidence in predicting a specific photon state, and the state with the highest score is determined as the outcome.
As the QLP represents the deviation of correlation results from the theoretical values of an ideal \(|n\rangle\) quantum emitter due to experimental factors, the exact value is not included in the machine learning outcome.
Instead, by discretizing QLPs into bins, the regression task is reformulated into a classification problem which aligns better with the CNN's specialty \cite{stewart_regression_2023}.

This model consists of 11 layers (excluding the input layer) and contains over 150,000 trainable parameters.
Except for the dense layers, batch normalization is applied to the output of each layer (not shown in the diagram).
The Adam method is chosen as the model optimizer and all CNN layers use ReLU (Rectified Linear Unit) as the activation function.
The \(g^{(3)}\) data is simulated for \(|1\rangle\), \(|2\rangle\) and \(|3\rangle\) Fock states, with two variable parameters: QLP and the number of detection events.
The QLP has 21 possible values ranging from 0 to 1 with intervals of 0.05, to represent the quantum pureness of the emitter.
Qualified quantum emitters require a QLP greater than or equal to 0.5, while a QLP less than 0.5 will be categorized as a coherent state.
The number of detection events that simulates the experiment time ranges from 100 to 100,000. 
Values from 100 to 10,000 are taken at intervals of 100, while values from 10,000 to 100,000 are taken at intervals of 1000, totaling 190 values.
A total of 11,970 cases are simulated, with 100 measurements for each case, resulting in 11,970,000 \(g^{(3)}\) correlation datasets.
70\% of the datasets are used for model training, with a batch size of 32 and a single epoch, given the substantial data that doesn't necessitate iterations.
20\% of the data are validation sets for hyperparameter tuning, while the remaining 10\% is used for testing the accuracy.
Considering the study's emphasis on improving fitting accuracy for sparse data, the model's evaluation primarily relies on the required number of photon detection events to achieve satisfactory predictions, instead of computational performance metrics such as computation time.

\section{Results}
\begin{figure}[!t]
    \includegraphics{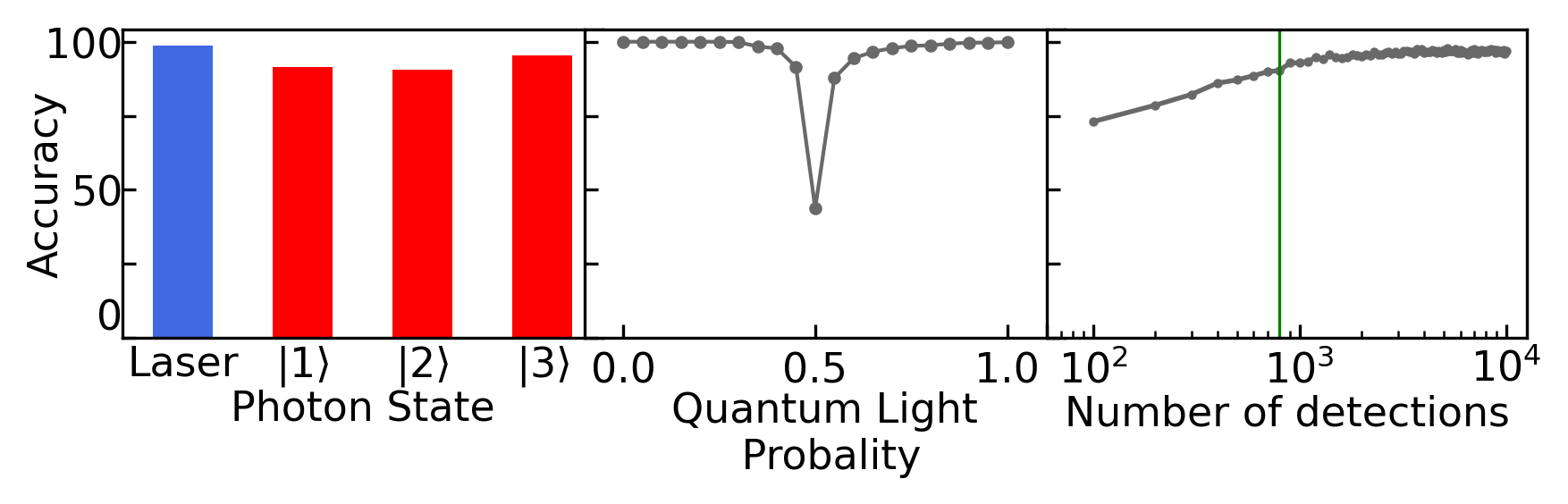}
    \caption{
    Accuracy of the CNN-based photon state classifier.  
    Left: Overall accuracy for each light source category. Blue: Laser with a coherent state. Red: Quantum light Fock states.  
    Middle: Averaged accuracy over all light source categories as a function of quantum light probability, which indicates the portion of quantum light in the simulation when it is mixed with a corresponding coherent state. A probability of 0.5 signifies a light source composed of equal parts quantum light and laser, while 0.0 or 1.0 indicates a pure laser or pure quantum light, respectively.  
    Right: Averaged accuracy over all light source categories as a function of number of detection events. The number of detections are in a logarithmic scale for better visualization. The green vertical line indicates the 90\% accuracy boundary. 
    }
\end{figure}
The average accuracy of classification results is shown in Figure 4, as a function of variables including the photon state, QLP, and the number of detection events.
The left bar chart displays the average accuracy for all cases in each photon state, with the coherent state plotted in blue, and the Fock states \(|1\rangle\), \(|2\rangle\) and \(|3\rangle\) are in red.
An accuracy of 90\% is achieved for all photon states, with classification of the coherent state and Fock state \(|3\rangle\) being higher compared to \(|1\rangle\) and  \(|2\rangle\).
The middle and right plots describe the accuracy over all the photon states as functions of QLP and the number of events.
A significant drop in accuracy is observed when the QLP approaches 0.5. 
This is due to the $QLP=0.5$ is defined as the threshold distinguishing between quantum light (Fock state) labels and non-quantum light (coherent state) labels.
Near this decision boundary, the similarity in the correlation results posts a challenge for accurate classification, which is further indicated in Fig.5.
As shown on the right, the average accuracy across all light categories improves from 72\% to 95\% as the number of simulation events rises from 100 to 10,000. 
The accuracy boundary of 90\%, achieved with only 800 events, is depicted by the green vertical line.

\begin{figure}[!t]
    \includegraphics{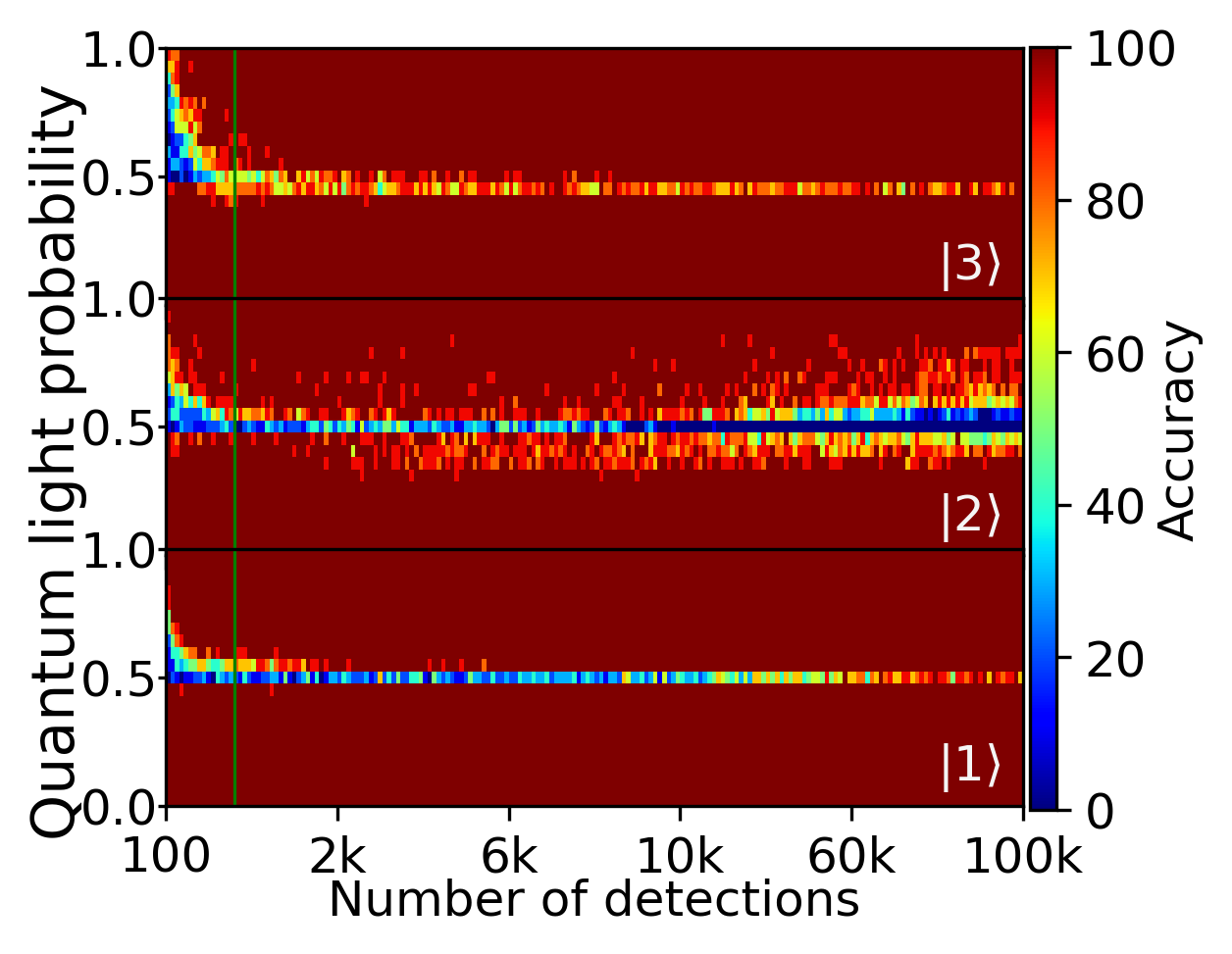}
    \caption{
    Accuracy distribution of photon Fock state classification, as a function of quantum light probability and number of detections for each Fock state case.  
    From top to bottom: Accuracy distributions for light sources, comprising $|3\rangle$, $|2\rangle$, or $|1\rangle$ Fock states of quantum light, respectively, each combined with its corresponding coherent state $|\alpha\rangle$, where the average photon number is 3, 2, or 1, respectively. For each case, a quantum light probability of 0.5 signifies a light source composed of equal parts quantum light and laser, while 0.0 or 1.0 indicates a pure laser or pure quantum light, respectively.  
    Green vertical line: boundary of 90\% averaged accuracy over all light source categories, at 800 detection events.
    }
\end{figure}
To explicitly illustrate the model's performance on a case-by-case basis, Fig.5 presents accuracy heatmap for the three Fock states, where each pixel is colored according to the average accuracy of all cases with a specific QLP and number of events.
The accuracy of coherent state classification is included in each Fock state where QLP is less than 0.5, instead of a separate plot.
In general, the accuracy is significantly enhanced with an increasing number of events across most QLPs.
A consistent accuracy drop occurs at the decision boundary with QLP approaching 0.5, which aligns with the trend in the middle figure of Fig.4.
For all three cases, the performance decreases at the left of the green line (the 90\% overall accuracy boundary from Fig.4, right), where datasets contain fewer detection events and have QLPs higher than 0.5.
While a relatively low accuracy seems reasonable with few-shot data, the "vertical" asymmetry of accuracy comes from the large data sparsity and uncertainty with higher QLPs.
In photon statistics, coherent states with larger variance are more likely to emit higher photon states \cite{fox_quantum_2006}, which produces recognizable features in the correlation.
For higher QLPs, the occurrence of coherent states drops down and consequently, leading to less distinctive features in the correlation results for recognition.
As the number of simulation events increases, the \(g^{(3)}\) results tend to stabilize for both coherent states and Fock states, which leads to quick disappearance of the triangular region with low accuracy.

For the \(|1\rangle\) case, the low-accuracy triangle vanishes earlier than \(|2\rangle\) and \(|3\rangle\) with growing detection events.
According to Table 1, the \(|1\rangle\) state exhibits larger variations in \(g^{(2)}(0)\) and \(g^{(3)}(0)\) values with changing QLP, which contributes greatly to the CNN classification.
Conversely, the smaller differences for \(|2\rangle\) and \(|3\rangle\) make them less conducive for recognition with fewer events, resulting in lower rate of increase in accuracy.
At the decision boundary with a QLP of 0.5, the accuracy of the \(|3\rangle\) state exhibits a higher rate of increase compared to \(|1\rangle\).
This is due to the relatively higher number of photons from the \(|3\rangle\) state facilitates pattern recognition by the model and consequently requires fewer events for accurate classification.

For the \(|2\rangle\) state, the accuracy shows a decrease with more detection events, particularly at intermediate QLPs.
Some mispredictions of coherent states can be attributed to the confusion near the decision boundary, as mentioned earlier, the majority of mispredictions involve the \(|3\rangle\) state, potentially caused by overfitting \cite{chauhan_convolutional_2018, alzubaidi_review_2021}.
Given that \(|2\rangle\) and \(|3\rangle\) exhibit similar values of \(g^{(2)}(0)\), the decision boundary \(g^{(2)}(0)=0.83\) (determined by Eqs. (1)) for distinguishing \(|3\rangle\) and its coherent state is misused for identifying \(|2\rangle\). 
Consequently, for the \(|2\rangle\) cases with \(g^{(2)}(0)\) greater than 0.83 (or the QLP exceeds 0.3), misclassification as \(|3\rangle\) results in a noticeable drop in accuracy, even with increased numbers of detection events.
While they possess different \(g^{(3)}(0)\) values, the contribution of this single element to the classifier is overshadowed by the abundance of \(g^{(2)}(0)\) elements in the correlation matrix.
Given that typically low photon co-detection events are involved in quantum correlation experiments \cite{thomay_simultaneous_2017a}, this paper primarily focuses on datasets containing fewer than 10,000 detection events.

\begin{figure}[!t]
    \includegraphics{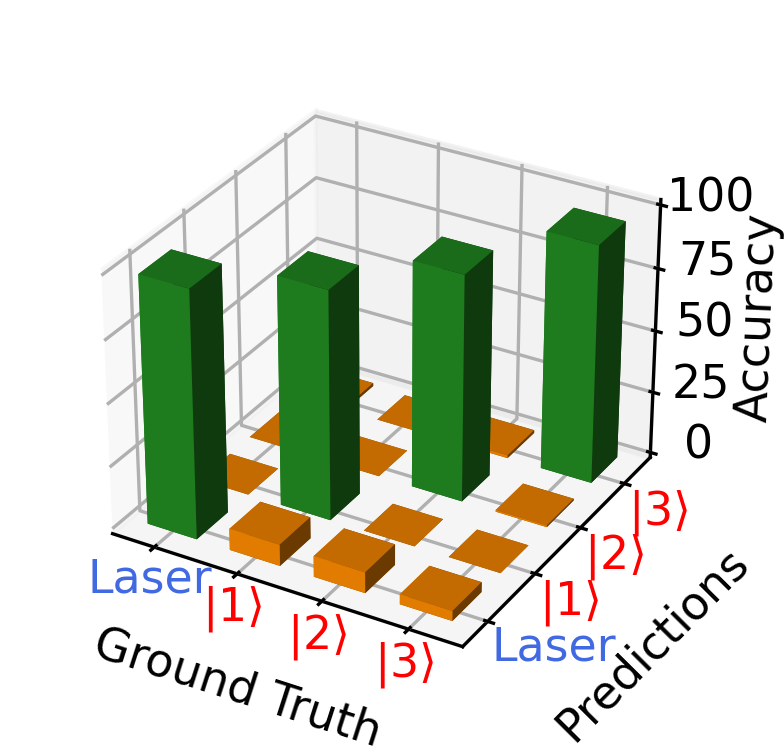}
    \caption{
    The confusion matrix of the classifier in a 3D bar plot. The correct predictions are represented by the diagonal elements in green, while the off-diagonal elements, indicating incorrect predictions, are depicted in yellow. Blue labels: Laser with a coherent state. Red labels: Quantum light Fock states.
    }
\end{figure}
To visualize the prediction distribution over photon states, a 3D bar chart of the confusion matrix is shown in Fig.6.
The diagonal elements in green indicate the average accuracy for each state, while the off-diagonal elements in yellow describe the distribution of misclassification.
The highest accuracy achieved is 98.7\% for the coherent state, while the accuracy for Fock states remain above 90\%.
This is attributed to more datasets are simulated for coherent states, which boosts the ML algorithm during the training process.
The model showcases a balanced performance in recognizing the Fock states, with the most common incorrect prediction being the coherent state, owing to their similarity near the decision boundary.
Another notable misclassification is observed between \(|2\rangle\) and \(|3\rangle\), which arises from the minor difference in the \(g^{(3)}(0)\) values as discussed earlier.
Other elements in the error matrix are below 0.1\% and can be considered negligible.

\section{Conclusion}
\begin{figure}[!t]
    \centering
    \includegraphics{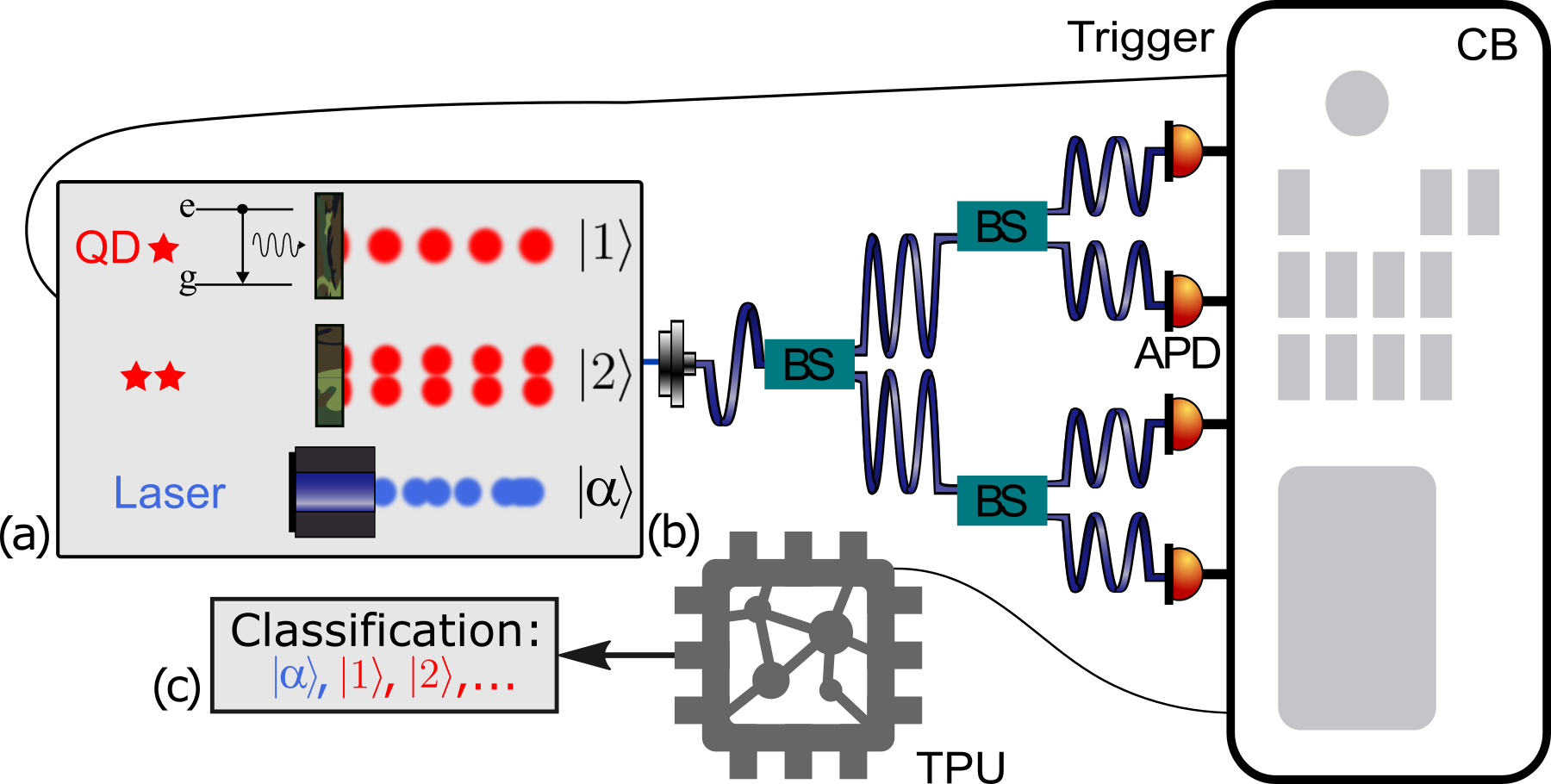}
    \caption{
    Experiment layout for measuring quasi real-time Fock state distribution enabled by Machine Learning.  
    (a) Light sources including: quantum light emitters, specifically two-level systems in a semiconductor quantum dot, with potential photon Fock states such as $|1\rangle$ or $|2\rangle$ (red dots); an attenuated laser in a coherent state $|\alpha\rangle$ (blue dots) with Poissonian photon distribution.  
    (b) Optical setup. The photon emission is coupled into an optical fiber, then equally split into four paths by the 50:50 Fiber Beam Splitter (BS). Each path leads to a Avalanche Photodiode (APD) that is connected to the Correlation Board (CB).   
    (c) Correlation data will be analyzed by an optimized CNN model in the Tensor Processing Units (TPU) that is a specialized hardware accelerator, which returns the predicted light source category.
    }
\end{figure}
In contrast to the fitting methods that can be implemented directly, the machine learning algorithm requires an additional training phase prior to data analysis.
However, this tradeoff can yield substantial enhancement of fitting accuracy and efficiency \cite{freire_artificial_2023}.
By achieving an average accuracy of 90\% with only 800 simulated events, the ML model demonstrates the capability for reliable classification without requiring lengthy measurement time.
Another advantage of ML algorithms is their ability to be optimized for recognizing specific data pattern \cite{jordan_machine_2015}.
As 2D CNNs have been extensively developed for recognizing spatial features \cite{lecun_convolutional_2010}, the integration of \(g^{(3)}\) with 2D CNN holds great potential for a novel quantum light classifier, especially in the multiphoton regime.

With Figure 7, we propose the implementation of the 2D CNN model in quasi real-time photon state distribution measurements.
Examples of different light emitters are included in Fig.7 (a).
The model on the top shows single-photon emission from the exciton recombination in a two-level system such as a semiconductor quantum dot. 
The red dots represent the single photons emitted periodically in the temporal regime, with the photon Fock state labeled as \(|1\rangle\).
The middle section illustrates a two-photon emitter that emits two indistinguishable photons represented by the red dots in pairs upon each excitation, with a Fock state of \(|2\rangle\).
The photon stream from a continuous-wave laser is shown at the bottom, where each emission occurs randomly in time, exhibiting a coherent photon state \(|\alpha\rangle\).
These listed light sources are coupled into an optical fiber system with an HBT configuration, indicated by the dark blue wavy lines in Fig.7 (b).
The incident beam path is evenly split into four output paths by the 50:50 beam splitters (BS).
Each output is attached to an avalanche photodiode (APD) that records the photon arrival time and transmits the data to the correlation electronics (CB).
\(g^{(3)}\) correlation data is computed online and fed into a Tensor Processing Unit (TPU), serving as a machine learning hardware accelerator.
The ML model is pre-trained and optimized to provide real-time photon state classification results, as shown in Fig.7 (c).
With the specialty of rapid fitting on sparse datasets, the measurement time on individual samples can be substantially reduced, and immediate feedback from the ML analysis enables real-time parameter optimization.

With the development of ML software \cite{abadi_tensorflow_, kerasteam_2024, paszke_pytorch_2019}, further enhancement on the presented prototype becomes possible.
For example, 2D locally-connected layers can be a promising substitute for the CNN layers for capturing the pattern (shown on Fig.2 left) that are spatially fixed in correlation data. 
In these layers, the parameter-sharing scheme between convolutional kernels is relaxed, which facilitates the recognition of specific structures \cite{bruna_spectral_2014}.
However, the model's capacity will be increased by incorporating multiple such layers.
Moreover, an ensemble of similar models can be developed, with each model trained for specific cases, to enhance the overall performance.

The presented simulation model can be tailored to simulate broader scenarios involving random coherent lights with low average photon numbers mixing with quantum emission. 
This adaptation is particularly relevant as quantum light detection often involves laser emission, as the excitation source of quantum emitters.
Additionally, this model can be further modified to compute dynamic correlations, such as the conditional auto-correlation function (CACF) for studying heralded photons \cite{shih_universal_2024}.
As the landscape of quantum emitters has been enriched by the emerging 2D materials with functional heterostructures and satisfactory fabrication efficiency such as Graphene \cite{zhao_single_2018}, TMD (Transition Metal Dichalcogenides) \cite{palacios-berraquero_largescale_2017, tran_quantum_2016, srivastava_optically_2015}, and Moire superlattices \cite{yu_moire_2017}, the scalability for photonic circuits is often impeded by the rare and random occurrence of quantum emitters \cite{palacios-berraquero_largescale_2017}, which are considered from defects or strains in these atomically thin layers \cite{he_single_2015, tonndorf_singlephoton_2015}.
This study offers insights into Fock state mapping on 2D materials, rapidly determining the spatial location, and assessing the quantum purity of potential quantum emitters.

Recently, novel methods for multiphoton preparation have been demonstrated.
This letter \cite{uria_deterministic_2020} proposed a theoretical model of deterministic generation of large Fock states up to \(|100\rangle\) based on the resonant interaction between a coherent state with two-level systems.
The experimental generation of eight indistinguishable photons using temporal-to-spatial demultiplexing with single photon sources was presented \cite{hansen_singleactiveelement_2023}.
The proposed \(g^{(3)}\) correlation - 2D CNN combination introduces new possibilities for categorizing and characterizing multiphoton states.

Additionally, this approach opens up the possibility for optimizing photosynthesis through real-time quantum feedback.
As the research \cite{rosenthal_biocompatible_2011} has shown that the implantation of biocompatible quantum dots (BQD) enables quantum measurements for monitoring intracellular environments, effectively overcoming the persistent challenge of strong autofluorescence background in traditional spectroscopy of plant cells \cite{cell_2005}.

In this letter, we present a machine learning model for quasi real-time categorization of photon states with \(g^{(3)}\) correlation, and propose the implementation in experiments.
This methodology introduces new feasible solutions for identifying quantum emitters and holds broad applications in quantum metrology within the field of nanophotonics.

%


\bibliography{Manuscript}

\end{document}